\documentclass[3p,times,twocolumn]{elsarticle}

\usepackage{ecrc}

\volume{273-275C}

\firstpage{1938}

\journalname{Nuclear Physics B Proceedings Supplement}

\runauth{B. A. Kniehl et al.}

\jid{nuphbp}

\jnltitlelogo{Nuclear Physics B Proceedings Supplement}

\usepackage{amssymb}
\usepackage[figuresright]{rotating}

\begin{document}

\begin{frontmatter}

%% Title, authors and addresses

%% use the tnoteref command within \title for footnotes;
%% use the tnotetext command for the associated footnote;
%% use the fnref command within \author or \address for footnotes;
%% use the fntext command for the associated footnote;
%% use the corref command within \author for corresponding author footnotes;
%% use the cortext command for the associated footnote;
%% use the ead command for the email address,
%% and the form \ead[url] for the home page:
%%
%% \title{Title\tnoteref{label1}}
%% \tnotetext[label1]{}
%% \author{Name\corref{cor1}\fnref{label2}}
%% \ead{email address}
%% \ead[url]{home page}
%% \fntext[label2]{}
%% \cortext[cor1]{}
%% \address{Address\fnref{label3}}
%% \fntext[label3]{}

\dochead{}
%% Use \dochead if there is an article header, e.g. \dochead{Short communication}

\title{Prompt-photon plus jet photoproduction with ZEUS at DESY HERA\\ in the parton Reggeization approach}

%% use optional labels to link authors explicitly to addresses:
%% \author[label1,label2]{<author name>}
%% \address[label1]{<address>}
%% \address[label2]{<address>}

\author[DESY]{B. A. Kniehl}
\author[SSAU]{M. A. Nefedov\fnref{speaker}}
\author[SSAU]{V. A. Saleev}

\tnotetext[speaker]{Speaker. E-mail addresses: {\tt kniehl@mail.desy.de}, {\tt nefedovma@gmail.com}, {\tt saleev@samsu.ru}}

\address[DESY]{{II.} Institut f\"ur Theoretische Physik, Universit\" at Hamburg,
Luruper Chaussee 149, 22761 Hamburg, Germany}
%\address[SSU]{Samara State University, Ac.\ Pavlov, 1, 443011 Samara, Russia}
\address[SSAU]{Samara National Research University, Moscow Highway, 34, 443086, Samara, Russia}
\begin{abstract}
%% Text of abstract
We study the photoproduction of isolated prompt photons associated with hadron
jets in the framework of the parton Reggeization approach.
The main improvements with respect to previous studies in the
$k_T$-factorization framework include the application of the Reggeized-quark
formalism, the generation of exactly gauge-invariant amplitudes with off-shell
initial-state quarks, and the exact treatment of the $\gamma R\to \gamma g$ box
contribution with off-shell initial-state gluons.
In this proceedings, the new data set, published recently by ZEUS collaboration is analyzed, were the distributions in photon and jet rapidity, transverse energy, azimuthal angle between photon and jet and proton momentum fraction are presented for different values of measured photon momentum fraction $x_\gamma^{meas} <0.7$, $0.8$ and $x_\gamma^{meas}>0.8$. The good agreement of measured distributions with our predictions is  observed for the direct-dominating part of the data set. The comparison with the previous calculations in $k_T$-factorization, role of nonfactorizable higher-order and hadronization corrections is discussed.
\end{abstract}

\begin{keyword}
%% keywords here, in the form: keyword \sep keyword

photoproduction \sep prompt-photons \sep jets \sep higher-order corrections in QCD \sep proton unintegrated PDFs \sep photon PDFs

%% MSC codes here, in the form: \MSC code \sep code
%% or \MSC[2008] code \sep code (2000 is the default)

\end{keyword}

\end{frontmatter}

%%
%% Start line numbering here if you want
%%
% \linenumbers

%% main text
\section{Introduction.}
The photoproduction of prompt photons with large transverse momenta provides a
formidable laboratory for precision tests of perturbative quantum
chromodynamics (QCD) and a useful source of information on the parton content
of the proton and the real photon.
The initial-state photon may interact with the partons inside the proton either
directly (direct photoproduction) or via its partonic content
(resolved photoproduction).

The inclusive photoproduction of prompt photons, singly and in association with
jets, received a lot of attention, both experimentally and theoretically.
On the experimental side, the H1~\cite{H1_data1,H1_data2} and
ZEUS~\cite{ZEUS_data1,ZEUS_data2,ZEUS_data3,ZEUS_data4} collaborations measured the
cross section distributions in the transverse energies $(E_T)$ and the
pseudorapidities $(\eta)$ of the prompt photon and the jet as well as in
azimuthal-decorrelation parameters such as the azimuthal angle enclosed between
the prompt-photon and jet transverse momenta ($\Delta\phi$) and the component
of the prompt-photon transverse momentum orthogonal to the direction of the jet
transverse momentum ($p_\perp$).
Also, the distributions in the variables estimating the momentum fractions of
the initial-state partons, $x^{\rm LO}_p$, $x^{\rm LO}_\gamma$, and
$x^{\rm obs}_\gamma$, were measured.
This rich set of observables allows one to perform a detailed study of the
underlying partonic processes and to assess the relevance of different
perturbative corrections.

On the theoretical side, attempts to describe this data where made both at
next-to-leading order (NLO) in the conventional collinear parton model (CPM)
\cite{NLO_FGH,NLO_KZ} and in approaches accommodating off-shell initial-state
partons, such as the $k_T$-factorization approach (KFA) \cite{LZ1,LZ2,MLZ} and $k_T$-factorization approach with Reggeized partons, which we refer to as the parton Reggeization approach (PRA) \cite{Sal_prompt_photon_HERA,KNS_photon_jet}.

For prompt-photon plus jet associated photoproduction, NLO CPM predictions
generally agree with the measured $\eta$ distributions, slightly underestimate
the $E_T$ distributions, and provide a poor description of the azimuthal
decorrelation observables \cite{H1_data2}, due to the fact, that these
distributions collapse to delta functions in the LO CPM and, therefore,
strongly depend on the radiation of additional partons.
The available KFA predictions provide a better description of the measured
$E_T$ distributions and azimuthal decorrelation observables, but are
implemented with matrix elements that manifestly violate gauge invariance,
which renders the quantitative improvements of the predictions questionable.
Furthermore, in the early studies \cite{LZ1,LZ2}, the partonic subprocess
pertaining to the scattering of a photon and an off-shell gluon,
$\gamma g^*\to \gamma g$, was not taken into account.
Later, this contribution was found to be numerically significant \cite{MLZ},
due to the large gluon luminosity under HERA conditions.
But the treatment of this contribution was approximate because the virtuality
of the initial-state gluon was not taken into account at the amplitude level,
but only in the kinematics of the process \cite{MLZ}.

In view of the shortcomings of the previous calculations mentioned above, the improved analysis of prompt-photon plus jet associated photoproduction in the LO of PRA was performed in \cite{KNS_photon_jet}. In this work, the contributions of the following of partonic subprocesses was taken into account:

 \begin{eqnarray}
  Q(q_1)+\gamma(q_2) &\to & q(q_3)+\gamma (q_4), \label{dir_compton} \\
  R(q_1)+\gamma(q_2) &\to & g(q_3)+\gamma (q_4), \label{dir_box} \\
  R(q_1)+q\left[\gamma\right](\tilde{q}_2) &\to & q(q_3)+\gamma(q_4), \label{res_compton1}
  \end{eqnarray}

  where the four-momenta of the partons are denoted in the brackets. The parton coming from the proton is taken to be off-shell ($q_1^2=-{\bf q}_{T1}^2=-t_1$), and carries one large light-cone component of momentum $q_1^+=2x_1 E_p\ll q_1^-$. This special (Multi-Regge) kinematics allows us to use the formalism of Reggeized gluons~\cite{LipatovEFT}, denoted by $R$ in (\ref{dir_compton}-\ref{res_compton1}), and quarks~\cite{LipVyaz}, denoted by $Q$, to define the gauge-invariant amplitude of the hard subprocess. See Ref.~\cite{KNS_photon_jet} for further explanations and references.

  The subprocess (\ref{dir_box}) is loop-induced, but due to the large gluon luminosity in the HERA kinematical conditions it's contribution is comparable to the contribution of the resolved subprocess (\ref{res_compton1}). The full $t_1$-dependence of the amplitude of the subprocess (\ref{dir_box}) was calculated in~\cite{KNS_photon_jet}, and the suppression of this contribution up to 30\% w. r. t. LO CPM result was observed as a consequence of this dependence.

  The contributions of other resolved subprocesses was found to be numerically negligible as well as the contribution of the parton to photon fragmentation, which is suppressed by the photon isolation condition, applied in the experimental analysis, see the discussion in~\cite{KNS_photon_jet} for the further details.

  In our numerical analysis we use the Kimber-Martin-Ryskin (KMR)~\cite{KMR} procedure to obtain the unintegrated PDF (unPDF) $\Phi_{i/p}(x_1,t_1,\mu_{Fp}^2)$ of the parton $i=g,q,\bar{q}$ in proton from the conventional (integrated) PDF of the CPM $f_{i/p}(x_1,\mu_{Fp})$. As a collinear input for the KMR procedure we have used the LO proton PDF set by Martin {\it et. al.}~\cite{MRST} with $n_F=4$ active quark flavors. To calculate the resolved contributions we have used the LO photon PDF set $f_{i/\gamma}(x_2,\mu_{F\gamma})$ by Gl\"uck {\it et al.}~\cite{GRV}.

  In this proceedings we compare our predictions for the prompt photon+jet associated photoproduction at HERA with recently published dataset~\cite{ZEUS_data4}, which we will refer to as ZEUS-2014.

\section{Numerical results for the ZEUS-2014 dataset.}

  The kinematic conditions for the ZEUS-2014 dataset are the following: the proton and electron energies where equal to $E_p=920$ GeV and $E_e=27.5$ GeV, and the photoproduction events are characterized by the range of inelasticity $0.2<y<0.7$ and invariant squared momentum transfer $Q^2<1$ GeV$^2$. The kinematical cuts applied on the photon and jet in the ZEUS-2013 dataset~\cite{ZEUS_data3} and ZEUS-2014 dataset~\cite{ZEUS_data4} are the same: 6.0~GeV${}<E_T^\gamma<15.0$~GeV, $-0.7<\eta^{\gamma}<0.9$, 4.0~GeV${}<E_T^{\rm jet}<35.0$~GeV, $-1.5<\eta^{\rm jet}<1.8$. But in the ZEUS-2014 dataset the kinematic distributions where presented in a different ranges of the measured photon momentum fraction transferred to the photon and the jet:
  \[
  x_\gamma^{meas}=\frac{E^\gamma+E^{jet}-p_Z^\gamma-p_Z^{jet}}{E^{all}-p_Z^{all}},
  \]
  where $all$ denotes all particles observed in the event. In the LO PRA, as well as in the LO CPM, the contributions of the direct subprocesses~(\ref{dir_compton},\ref{dir_box}) to the cross section is proportional to $\delta(x_\gamma^{meas}-1)$, because the Reggeized parton do not carry the $q_1^-$ momentum component. Resolved contributions lead to the non-trivial dependence on the $x_\gamma$ already in the LO. In the Ref.~\cite{ZEUS_data4}, the cross sections differential in $\eta^\gamma$, $\eta^{jet}$, $E_T^\gamma$, $E_T^{jet}$, $x_p$ and $\Delta\phi$ are presented for the direct-dominated region $x_\gamma^{meas}>0.8$ as well as for resolved-dominated regions $x_\gamma^{meas}<0.8$ and $x_\gamma^{meas}<0.7$, which allows one to perform the very detailed tests of the quality of the model.

    As it was shown in~\cite{KNS_photon_jet}, the dependence of the cross section on the variable $x^\gamma_{obs}$, defined in this paper, is well reproduced by our model for $x^\gamma_{obs}\leq 0.6$. The region $x^\gamma_{obs}>0.6$ seems to be direct-dominated, and the shape of the $x^\gamma_{obs}$-distribution is not reproduced there in the LO PRA (see Fig. 13 ~\cite{KNS_photon_jet}). To smear the direct-$x^\gamma_{obs}$ distribution, the NLO $2\to 3$ processes with the production of the additional parton in the central region of rapidity should be included.

    The authors of Ref.~\cite{MLZ} claim to do this, but the treatment of the double counting of the additional radiation between different perturbative orders of the hard process and between the hard process and unPDF is a serious issue in $k_T$-factorization. Including the $2\to 3$ processes they are forced to throw away their analogues of the LO $2\to 2$ processes (\ref{dir_compton}) and (\ref{res_compton1}), and the question of the double counting with the unPDF is not addressed in the Ref~\cite{MLZ} at all. As a resut, they reproduce well the $x^\gamma_{obs}$ distribution of Ref.~\cite{ZEUS_data3}, but fail to reproduce the shape of $\eta^{jet}$ distribution for $x_\gamma^{meas}<0.8$ (see Fig. 4(b) of the Ref.~\cite{ZEUS_data4}), which is well described in the NLO CPM and LO PRA, as it will be shown below.

    \begin{figure}
    \includegraphics[scale=0.435]{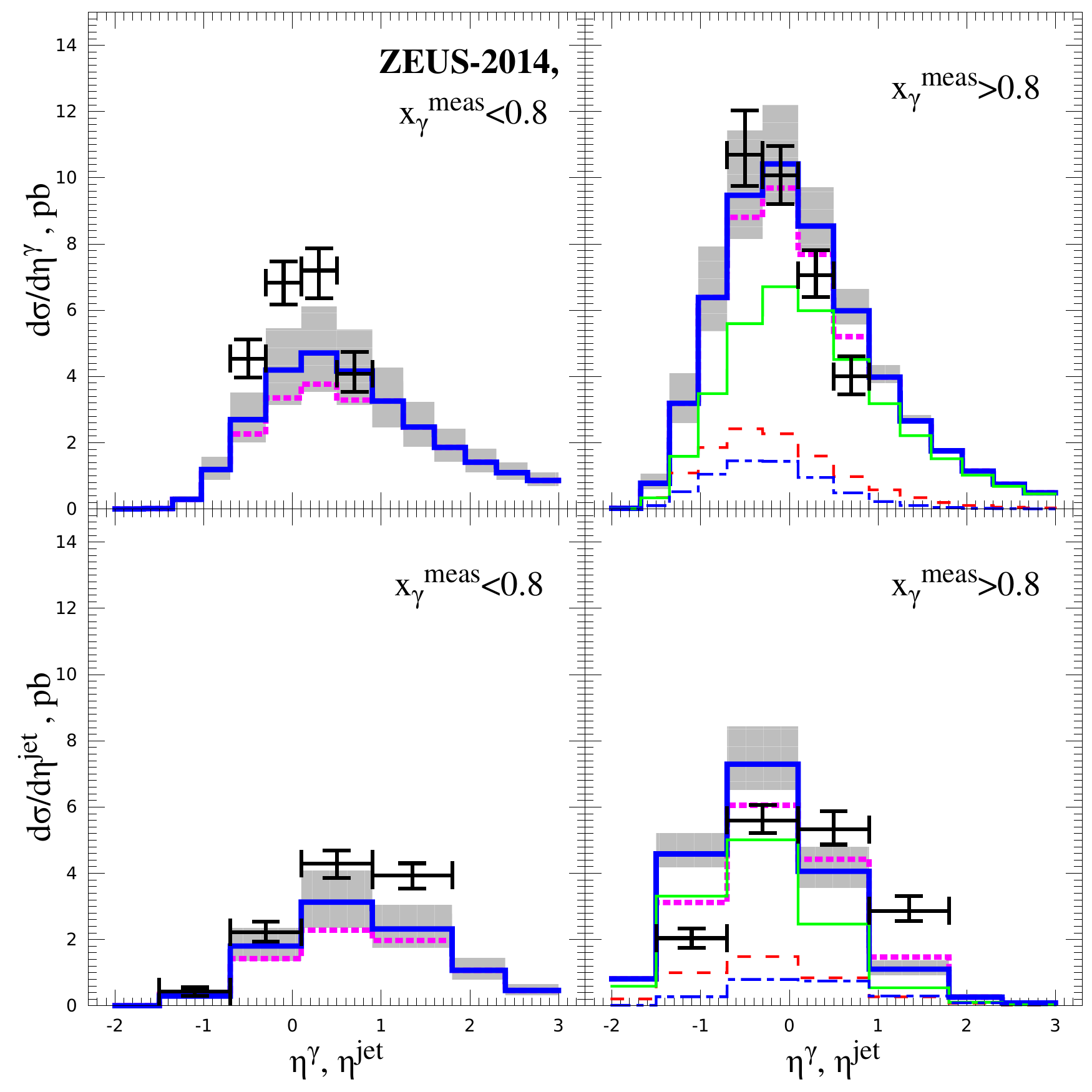}
    \caption{$\eta^\gamma$ and $\eta^{jet}$ distributions of $pe\to\gamma+j+X$ under ZEUS-2014 \cite{ZEUS_data4} kinematic conditions.
The experimental data are compared with LO PRA predictions at the parton level (boldfaced solid blue lines and the grey scale-uncertainty band) and with the hadronization corrections of the Ref.~\cite{ZEUS_data4} applied (boldfaced dotted magenta lines).
The LO PRA predictions are decomposed into the contributions due to the
subprocesses in Eqs.~(\ref{dir_compton}) (solid green lines), (\ref{dir_box})
(dashed red lines), and (\ref{res_compton1}) (dot-dashed blue lines), and only the last one contributes for $x_\gamma^{meas}<0.8$.\label{plot_eta}}
    \end{figure}

    \begin{figure}
     \includegraphics[scale=0.435]{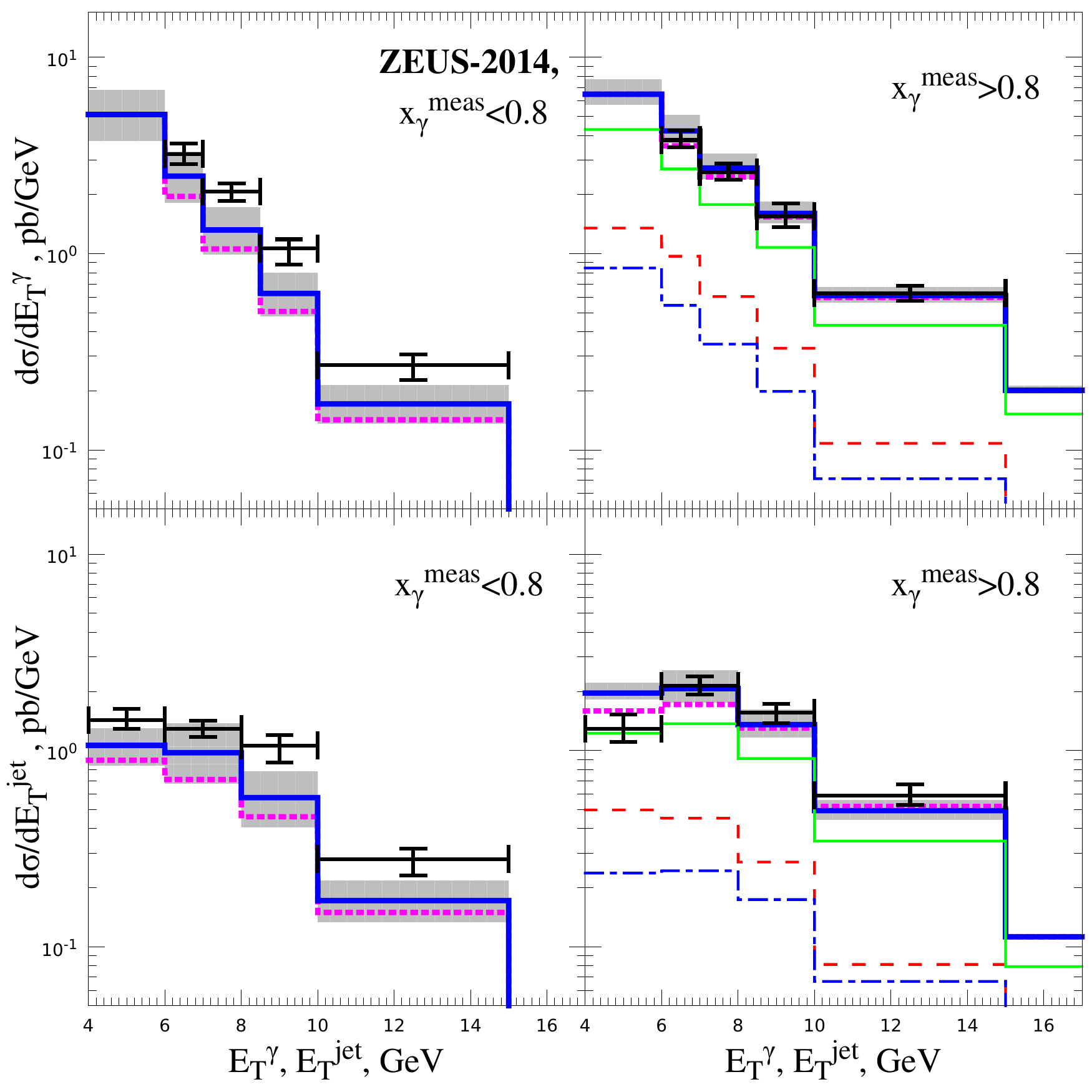}
    \caption{$E_T^\gamma$ and $E_T^{jet}$ distributions of $pe\to\gamma+j+X$ under ZEUS-2014 \cite{ZEUS_data4} kinematic conditions. The notations on the plots are the same as on the Fig.~\ref{plot_eta}.\label{plot_ET}}
    \end{figure}

\begin{figure}
    \includegraphics[scale=0.435]{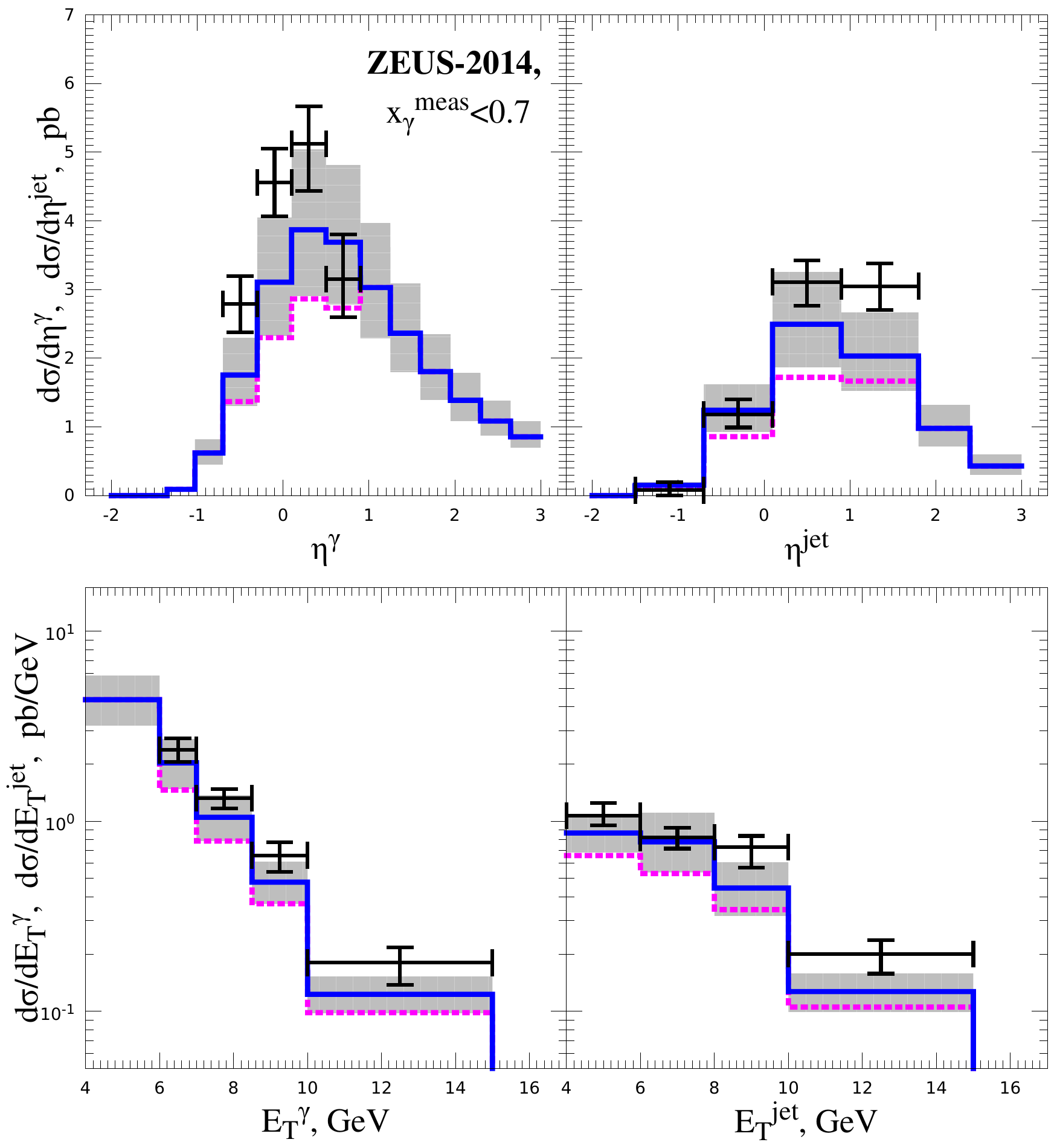}
    \caption{$\eta^\gamma$, $\eta^{jet}$, $E_T^\gamma$, $E_T^{jet}$ distributions of $pe\to\gamma+j+X$ under ZEUS-2014 \cite{ZEUS_data4} kinematic conditions for $x_\gamma^{meas}<0.7$. The notations on the plots are the same as on the Fig.~\ref{plot_eta}.\label{plot_low_xgamma}}
    \end{figure}

\begin{figure}
    \includegraphics[scale=0.435]{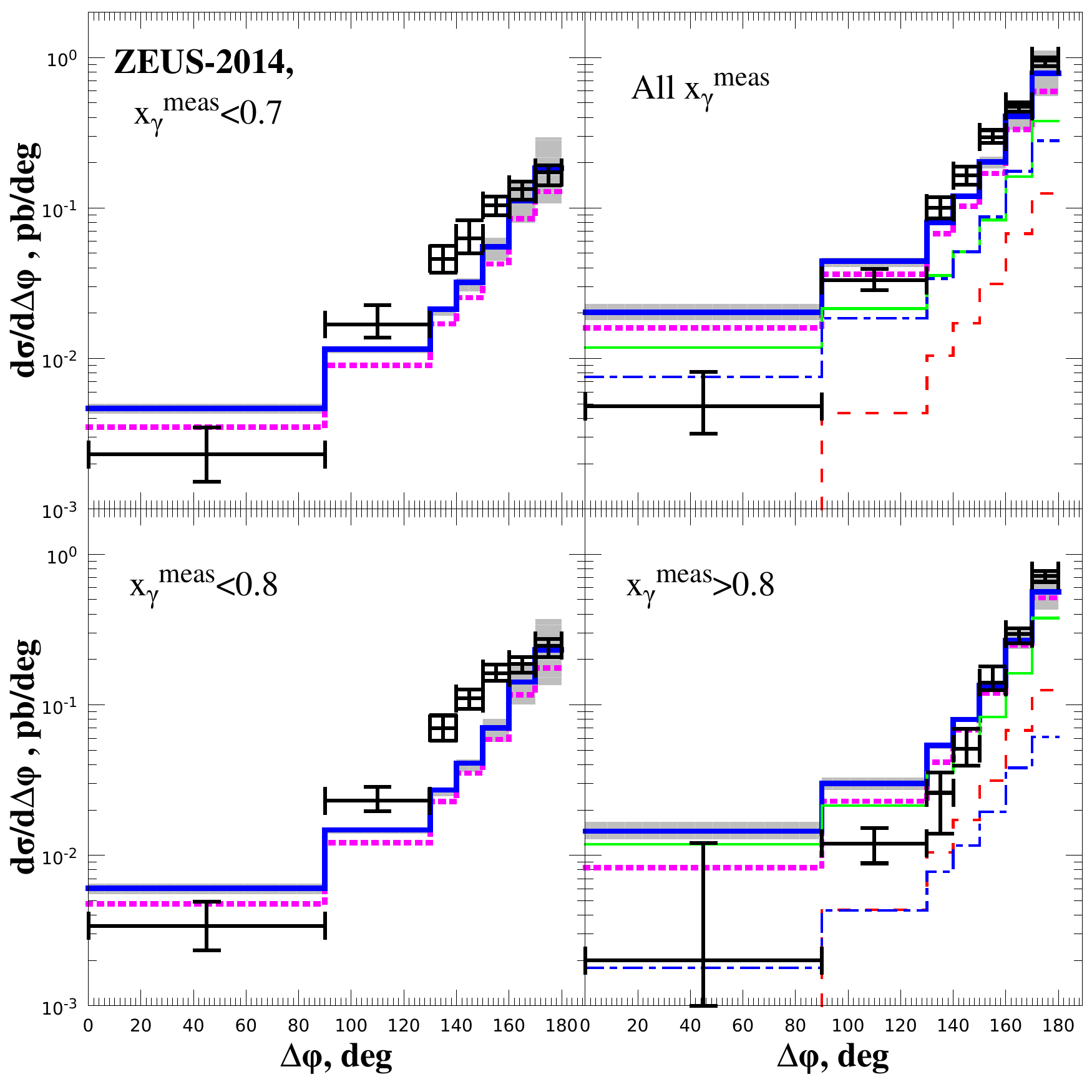}
    \caption{$\Delta\phi$ distributions of $pe\to\gamma+j+X$ under ZEUS-2014 \cite{ZEUS_data4} kinematic conditions. The notations on the plots are the same as on the Fig.~\ref{plot_eta}.\label{plot_Dphi}}
    \end{figure}

    In the figures \ref{plot_eta} -- \ref{plot_Dphi} the predictions of our model for the cross sections differential in $\eta^\gamma$, $\eta^{jet}$, $E_T^\gamma$, $E_T^{jet}$ and $\Delta\phi$ are shown for $x_\gamma^{meas}<0.8$($<0.7$) and $x_\gamma^{meas}>0.8$, also on the Fig. \ref{plot_Dphi}, the differential cross section $d\sigma/d\Delta\phi$ for all values of $x_\gamma^{meas}$ is presented.

   The theoretical uncertainty shown in the figures is only due to variation of the renormalization scale $\mu_R$ and two factorization scales $\mu_{Fp}$ and $\mu_{F\gamma}$ in $2^{\pm 1}$ times around their common central value $\max(E_T^\gamma,E_T^{jet})$. In the present analysis we have varied each of this three scales separately, keeping the other two fixed, and we have taken the largest variation of the cross section in the each bin as an estimate for the uncertainty.

   Also we have studied the effect of hadronization corrections, which where calculated in Ref.~\cite{ZEUS_data4} as a ratio of the detector-level and parton-level differential cross sections obtained from PYTHIA Monte-Carlo Event Generator. These corrections where applied in~\cite{ZEUS_data4} both to the NLO CPM and $k_T$-factorization theoretical predictions.

   In the case of our model, the application of the hadronization corrections improves the description of the data for the direct-dominated part of the phase-space, especially for the $\eta^{jet}$ distribution of the bottom-right panel of the Fig.~\ref{plot_eta}. Hadronization corrections to this distribution are significant, and change the form of the distribution, improving the agreement with data. On the contrary, for the $x_\gamma^{meas}<0.8$ part of the data, hadronization corrections lead to the systematic underestimation of the data. This is probably a consequence of the fact, that the variable $x_\gamma^{meas}$ by construction is very sensitive to the additional(subleading) hard and soft radiation in the event, which is not taken into account in our LO computation. The variable $x_\gamma^{LO}$, used by H1 collaboration in~\cite{H1_data1,H1_data2}, and the variable $x_\gamma^{obs}$ used by ZEUS in the analysis~\cite{ZEUS_data3} depends only on the photon and jet momenta, so they should be less sensitive to the soft-radiation/hadronization effects, and the latter one is as good in separating between direct and resolved contributions as $x_\gamma^{meas}$, since in the LO of PRA and CPM the direct contributions are proportional to the $\delta(x_\gamma^{obs}-1)$.

   As it was stated above, we reproduce the shape of $\eta^{jet}$ distribution (bottom-left panel of the Fig.~\ref{plot_eta}) rather well already at the LO. The $\eta^\gamma$ distribution is underestimated, as well as $E_T^\gamma$, $E_T^{jet}$ (Fig.~\ref{plot_ET}) and $\Delta\phi$ (Fig.~\ref{plot_Dphi}) distributions for the $x_\gamma^{meas}<0.8$, probably due to the lack of direct contribution in this region.

   In the resolved-dominating region $x_\gamma^{meas}<0.7$ we reproduce the $\eta^\gamma$, $\eta^{jet}$, $E_T^\gamma$ and $E_T^{jet}$ rather well within experimental and theoretical uncertainties (Fig.~\ref{plot_low_xgamma}). The $\Delta\phi$ distribution for the $x_\gamma^{meas}<0.7$ (Fig.~\ref{plot_Dphi}, top-left panel) is also reproduced slightly better than for $x_\gamma^{meas}<0.8$ (Fig.~\ref{plot_Dphi}, bottom-left panel). This probably shows, that we have found a good LO approximation both for the direct and resolved subprocesses in the framework of $k_T$-factorization.

   All the distributions in the direct-dominated region $x_\gamma^{meas}>0.8$ are described on a same level of quality as in the Ref.~\cite{KNS_photon_jet}. As it can be observed form the right panels of the Fig.~\ref{plot_Dphi}, the shape of the $\Delta\phi$ spectra in the region $130^\circ<\Delta\phi<180^\circ$ is well reproduced for the $x_\gamma^{meas}>0.8$ and also for the all-$x_\gamma^{meas}$ spectra. The same result was obtained in the Ref.~\cite{KNS_photon_jet} for the normalized $\Delta\phi$ distributions in H1-2010 kinematics, but for ZEUS-2014 data, we are also able to reproduce the normalization of the cross section.

\section{Conclusions.}

  In this proceedings contribution, we have compared theoretical predictions of our LO PRA model of Ref.~\cite{KNS_photon_jet} with the recent experimental data~\cite{ZEUS_data4} on the prompt photon associated with jet photoproduction, meashured by the ZEUS Collaboration at DESY HERA. The reasonable quantitative agreement of our predictions with experiment in the direct-dominated part of the dataset, as well as a good qualitative agreement in with the resolved-dominated data provides a new test of $k_T$-factorization framework with Reggeized quarks and gluons.

%% The Appendices part is started with the command \appendix;
%% appendix sections are then done as normal sections
%% \appendix

%% \section{}
%% \label{}

%% References
%%
%% Following citation commands can be used in the body text:
%% Usage of \cite is as follows:
%%   \cite{key}         ==>>  [#]
%%   \cite[chap. 2]{key} ==>> [#, chap. 2]
%%

%% References with BibTeX database:
\nocite{*}
\bibliographystyle{elsarticle-num}
\bibliography{martin}

%% Authors are advised to use a BibTeX database file for their reference list.
%% The provided style file elsarticle-num.bst formats references in the required Procedia style

%% For references without a BibTeX database:

\end{document}